# A New Perspective of Graph Data and A Generic and Efficient Method for Large Scale Graph Data Traversal

Chenglong Zhang

**Abstract**—The BFS algorithm is a basic graph data processing algorithm and many other graph data processing algorithms have similar architectural features with BFS algorithm and can be built on the basis of BFS algorithm model. We analyze the differences between graph algorithms and traditional high-performance algorithms in detail, propose a new way of classifying algorithms into data independent algorithm and data correlation algorithm based on their run-time correlation with data, and use this new classification to explain the validity of the methods proposed in this paper. Through a deeper analysis of graph data, we propose a new fundamental perspective on understanding graph data, establishing a link between two basic graph structures, graph and tree, and viewing graph data as consisting of smaller subgraphs and edge trees. Small degree vertices are found to be one of important cause of random memory access. Based on this, we propose a general, easy to implement, and efficient method for graph data processing, with the basic idea of treating low-degree vertices and core subgraphs separately, thus significantly reducing the size of random memory access and improving the efficiency of memory access. Finally, we evaluated the performance of the method on three major data center computing platforms (Intel, AMD, and ARM), and the experiments showed that it brought 19.7%, 31.8% and 17.9% performance improvement, respectively, with a performance-power ratio of 282.70 MTEPS/s on the ARM platform, ranking it among the Green graph500 in November 2019. World No. 1 on the big dataset list.

**Index Terms**—Parallel algorithms, Breadth first search, Graph algorithms, Graph and tree search strategies, Graph500

✦

## 1 INTRODUCTION

DATA can be divided into structured data and unstructured data. Unstructured data is more difficult for a computer to understand as compared to structured data. Graph data is a typical example of unstructured data. Graph is highly abstract and flexible, and can adequately express the connections and dependencies of things in nature. Many problems can be solved efficiently with graph-related algorithms supported by graph theory, such as graph coloring, network routing, and network flow. In addition, graph data processing allows mining and analysis of huge, sparse, and ultra-dimensional associations, and has been widely used in social networks, transportation networks, bioinformatic networks, knowledge graphs, GNN, etc.[1],[2],[3],[4]. However, the scale of graph data increases exponentially, and the number of edges can reach billions, in addition, natural graphs often exhibit a very skewed power-law distribution [5], which brings a great challenge to computing systems at all levels, and how to handle large-scale graph data efficiently has become the focus of research in academia and industry.

The Breadth First Search (BFS) algorithm is a basic graph data processing algorithm, and many graph algorithms can be built based on the BFS algorithm, such as

PageRank, Single-Source Shortest Path, Connected Component, Betweenness Centrality, etc. [6]. Many mainstream graph computing frameworks and programming models are now extended into generic forms based on the BFS algorithm model, such as ligra [6], ligra+ [7], Gemini [8], and Grazelle [9]. In addition, numerous other graph processing algorithms are essentially the same as BFS algorithms in terms of their architectural features. These algorithms have significantly different architectural characteristics from those of traditional algorithms for high-performance computational processing (matrix multiplication, FFT, convolution, etc.). Graph processing algorithms have typical characteristics such as poor data locality, low memory access efficiency, low parallelism and poor scalability, and the processing of graph data reflects significant inefficiencies on high-performance computers, with design challenges at all levels of the computer[10]. Top500 ranking is used internationally to measure the performance and power consumption of computers and clusters, profoundly affecting the development of computers at all levels. The benchmark Top500 used is the traditional vector and matrix multiplication and other high-performance numerical algorithms, but there are some drawbacks in using such algorithms to measure computer efficiency. Therefore, the Graph500 ranking was proposed internationally in 2010 to evaluate the performance and power consumption of computers and clusters [11], which uses the BFS algorithm as the benchmark. In summary, if a basic algorithm like BFS can be studied in depth, it will facilitate the research of general graph com-puting frameworks, improve the performance of other graph processing algorithms, or indirectly provide new research ideas to the research of graph processing frameworks and graph algorithms. It will also promote the development of computers at all levels like Top500.

---

• This work has been submitted to the IEEE for possible publication. Copyright may be transferred without notice, after which this version may no longer be accessible.
• The authors are with the State Key Laboratory of Computer Architecture, Institute of Computing Technology, Chinese Academy of Sciences, School of Computer and Control Engineering，University of Chinese Academy of Sciences,Beijing 100049, China. E-mail: {zhangchenglong l@ict.ac.cn.

In this paper, the optimization technique of BFS algorithm under a single node will be systematically introduced. However, these methods treat all vertices uniformly and do not explore the specificity of low-degree vertices, resulting in low-degree vertices bringing a large number of random access to affect performance. We propose a new fundamental way to understand graph data and a fundamental BFS algorithm model to divide the low-degree vertices and core subgraphs to significantly reduce the random access size and improve the traversal efficiency of the graph processing problem. This optimization idea can be implemented both in a generic graph processing framework and on different platforms such as CPU/CPU cluster/GPU/ASIC. The main contributions of this paper are as follows:

- We propose a new algorithm classification approach by analyzing the differences between graph algorithms and high-performance numerical algorithms. And the effectiveness of the methods proposed in this paper is explained using this classification approach.

- We propose a new fundamental perspective of understanding graph data, which can be seen as smaller core subgraphs and edge trees.

- We find that small degree vertices are one of the most important reasons for the high random access of the BFS algorithm(degree 1, etc.), which are typically located on edge trees.

- We propose a new general, easy-to-implement, efficient and large graph data traversal method. The central idea of the method is to treat low-degree vertices and core subgraphs separately. The method improves the performance significantly while maintaining the generalization of the BFS algorithm pattern to build a graph processing framework.

- The method was fully performance evaluated on different computing platforms (Intel, AMD, ARM) and the results show that the method can significantly improve graph processing performance on different platforms. And it achieved the No.1 in the world in the Green graph500 large dataset list in November 2019.

The chapters of this paper are organized as follows. Chapter 2 systematically summarizes the common optimization methods of the BFS algorithm, which are also implemented in this paper. Chapter 3 introduces our proposed algorithm classification approach. Chapter 4 introduces a new fundamental perspective of understanding graph data. Chapter 5 introduces a strong and effective graph data traversal method. Chapter 6 presents the experimental results. Chapter 7 summarizes the full text.

## 2 RELATED WORK

### 2.1 Hybrid Optimized

The traditional BFS algorithm is a top-down traversal method that generates a large number of invalid detections later in the traversal. Beamer creatively proposed a Bottom-up algorithm to reduce invalid traversals [12]. As shown in Algorithm 1 , the Bottom-up algorithm uses the exact opposite idea to Top-down. It checks whether there are any neighbor vertices in the unvisited vertices that are located in the current layer, and if so, it breaks out of the loop and ends the access to the remaining neighbor vertices, effectively reducing the redundant access overhead. However, the Bottom-up algorithm generates a large number of invalid detections in the previous layers. By combining Top-down and Bottom-up, using Top-down in the early part of the traversal, and switching to Bottom-up in the middle and late part of the traversal, the traversal efficiency can be significantly improved.

---

**Algorithm 1** Direction-optimizing BFS

```
 1: Input
 2:    G = (V, A^{OUT}, A^{IN})    ▷ A^{OUT} represents the set of outgoing edges of
       the undirected graph, and A^{IN} represents the set of incoming edges of the
       undirected graph
 3:    vertex_start
 4: Output
 5:    parent[N]                    ▷ the parent info of every vertex
 6:
 7: parent[v] ← ⊥, ∀v ∈ V\{vertex_start}
 8: parent[vertex_start] ← vertex_start
 9: VS ← {vertex_start}            ▷ store the visit state of all vertices
10: CQ ← {vertex_start}            ▷ store the vertices in current level
11: NQ ← ∅                         ▷ store the vertices in next level
12:
13: while CQ ≠ ∅ do
14:    if is_top_down_direction(CQ, NQ, VS) then
15:       NQ ← TOP-DOWN(G,CQ,VS,parent)
16:    else
17:       NQ ← BOTTOM-UP(G,CQ,VS,parent)
18:    end if
19:    swap(CQ,NQ)
20: end while
21: return parent[N]
22:
23: procedure TOP-DOWN(G, CQ, VS, parent)
24:    NQ ← ∅
25:    for v ∈ CQ in parallel do
26:       for w ∈ A^{OUT}(v) do
27:          if w ∉ VS atomic then
28:             parent[w] ← v
29:             VS ← VS ∪ {w}
30:             NQ ← NQ ∪ {w}
31:          end if
32:       end for
33:    end for
34:    return NQ
35: end procedure
36:
37: procedure BOTTOM-UP(G, CQ, VS, parent)
38:    NQ ← ∅
39:    for w ∈ V\VS in parallel do
40:       for v ∈ A^{IN}(w) do
41:          if v ∈ CQ then
42:             parent[w] ← v
43:             VS ← VS ∪ {w}
44:             NQ ← NQ ∪ {w}
45:             break
46:          end if
47:       end for
48:    end for
49:    return NQ
50: end procedure
```

---

### 2.2 NUMA Optimized

With the increase of the number of cores in a processor, as well as the number of sockets, the single-chip memory interconnect architecture has become a bottleneck, so the NUMA architecture has developed into the dominant architecture. Yasui et al [13] proposed a NUMA graph partitioning method for this feature of the NUMA architecture, which preprocesses the NUMA data based on the features of the top-down and bottom-up algorithms, re-spectively. The method significantly improves the locality of NUMA access to graph data. Equation (1) denotes the set of vertices to which the kth NUMA is divided, where l denotes the number of numa nodes, n denotes the number of all vertices. Equation (2) denotes the adjacency list of out-edge neighbors to which the kth NUMA is assigned in the top-down algorithm, and equation (3) denotes the adjacency list of in-edge neighbors to which the kth NUMA is assigned in the bottom-up algorithm. In order to improve the overall numa locality, the current layer vertices CQ, visit information VS, the next layer vertices NQ, and parent array are also numa data partitioned. Algorithm 2 to perform the above NUMA division, top-down and bottom-up algorithm. The method significantly improves the NUMA locality of the graph data.

$$V_k = \{V_j \in V \mid j \in [\frac{k}{l} \cdot n, \frac{(k+1)}{l} \cdot n)\} \qquad (1)$$

$$A_k^{OUT}(v) = \{w \mid w \in \{V_k \cap A(v)\}\}, v \in V \qquad (2)$$

$$A_k^{IN}(w) = \{v \mid v \in A(w)\}, w \in V_k \qquad (3)$$

---

**Algorithm 2** NUMA-optimized BFS

1: **Input**
2:    $k \in \{0, 1, ..., l-1\}$        ▷ NUMA node indices
3:    $T = \{T_k\}$        ▷ NUMA local threads
4:    $G = \{G_k\} = \{V_k, A_k^{OUT}, A_k^{IN}\}$
5:    $CQ = \{CQ_k\}$        ▷ NUMA local frontier queues
6:    $VS = \{VS_k\}$        ▷ the visit state of vertices
7:    $parent = \{parent_k\}$        ▷ the parent info of every vertex
8: **Output**
9:    $NQ = \{NQ_k\}$        ▷ NUMA local next frontier queues
10: **procedure** NUMA-OPTIMIZED-TOP-DOWN($G, CQ, VS, parent$)
11:    $NQ_k \leftarrow \emptyset$
12:    **for** $v \in CQ_k$ **in parallel** $(T_k)$ **do**
13:      **for** $w \in A_k^{OUT}(v)$ **do**
14:        **if** $w \notin VS_k$ **atomic then**
15:          $parent_k(w) \leftarrow v$
16:          $VS_k \leftarrow VS_k \cup \{w\}$
17:          $NQ_k \leftarrow NQ_k \cup \{w\}$
18:        **end if**
19:      **end for**
20:    **end for**
21:    **return** $NQ_k$
22: **end procedure**
23:
24: **procedure** NUMA-OPTIMIZED-BOTTOM-UP($G, CQ, VS, parent$)
25:    $NQ_k \leftarrow \emptyset$
26:    **for** $w \in V_k \backslash VS_k$ **in parallel**($T_k$) **do**
27:      **for** $v \in A_k^{IN}(w)$ **do**
28:        **if** $v \in CQ_k$ **then**
29:          $parent_k(w) \leftarrow v$
30:          $VS_k \leftarrow VS_k \cup \{w\}$
31:          $NQ_k \leftarrow NQ_k \cup \{w\}$
32:          **break**
33:        **end if**
34:      **end for**
35:    **end for**
36:    **return** $NQ_k$
37: **end procedure**

---

## 2.3 Degree Aware Bottom Up

The Bottom-up algorithm scans all the neighbor vertices that have not been visited in its traversal, and ends the traversal of the remaining neighbor vertices as soon as a neighbor vertex is found in the current layer. The earlier it is terminated, the number of neighbor checks can be reduced. Yasui et al [13] experimentally found a correlation between vertex degree and access frequency, the higher the degree of the vertex, the higher its access frequency. As shown in Algorithm 3, splitting the neighborhood adjacency list of each vertex into $A^{IN+}$ containing only the highest in-degree neighbor vertices and the re-maining in-neighborhood list $A^{IN-}$ arranged in descending order by degree, and splitting the Bottom up algorithm into the processing of both adjacency lists, a large number of vertices will not only be successfully detected in $A^{IN+}$, but also $A^{IN+}$ is visited sequentially, lines 12-19 of the algorithm. This not only greatly reduces the traversal of redundant edges, but also improves the locality of data access.

---

**Algorithm 3** Degree-aware Bottom-up BFS

1: **Input**
2:    $k \in \{0, 1, ..., l-1\}$        ▷ NUMA node indices
3:    $T = \{T_k\}$        ▷ NUMA local threads
4:    $G = \{G_k\} = \{V_k, A_k^{IN+}, A_k^{IN-}\}$
5:    $CQ = \{CQ_k\}$        ▷ NUMA local frontier queues
6:    $VS = \{VS_k\}$        ▷ the visit state of vertices
7:    $parent = \{parent_k\}$        ▷ the parent info of every vertex
8: **Output**
9:    $NQ = \{NQ_k\}$        ▷ NUMA local next frontier queues
10: **procedure** DEGREE-AWARE BOTTOM-UP($G, CQ, VS, parent$)
11:    $NQ_k \leftarrow \emptyset$
12:    **for** $w \in V_k \backslash VS_k$ **in parallel**($T_k$) **do**
13:      $v \leftarrow A_k^{IN+}(w)$
14:      **if** $v \in CQ_k$ **then**
15:        $parent_k(w) \leftarrow v$
16:        $VS_k \leftarrow VS_k \cup \{w\}$
17:        $NQ_k \leftarrow NQ_k \cup \{w\}$
18:      **end if**
19:    **end for**
20:
21:    **for** $w \in V_k \backslash VS_k$ **in parallel**($T_k$) **do**
22:      **for** $v \in A_k^{IN-}(w)$ **do**
23:        **if** $v \in CQ_k$ **then**
24:          $parent_k(w) \leftarrow v$
25:          $VS_k \leftarrow VS_k \cup \{w\}$
26:          $NQ_k \leftarrow NQ_k \cup \{w\}$
27:          **break**
28:        **end if**
29:      **end for**
30:    **end for**
31:    **return** $NQ_k$
32: **end procedure**

---

## 2.4 Static Round-Robin Shuffle

The natural graph is a power-law graph with extremely unbalanced distribution of degrees and numbers of vertices, which leads to unbalanced multicore load and inefficient thread-level parallelism, and how to fully exploit the advantages of the multicore architecture becomes a fundamental problem. We propose a static round-robin shuffle optimization method that allocates vertices according to their degree by round robin[14]. As shown in Algorithm 4,The vertices are sorted in descending order of degree, and the vertices are assigned to different concurrent entities(node, numa, thread, etc.) by polling to ensure that the high degree vertices and low degree vertices are evenly assigned to each concurrent entities, and the vertices are still kept in descending order of degree in each concurrent entities. In practice, if numa data partitioning is used, then consider numa-level static round-robin shuffle data partitioning first, followed by thread-level data partitioning. By the above method, on one hand, the data locality of vertex ordering is maintained. On the other hand, it improves the sequential memory access to vertices in each thread. The overhead caused by frequent dynamic scheduling of threads and the empirical parameter adjustment of block granularity in the dynamic allocation method are avoided. When optimizing for a problem, if it is found that data preprocessing leads to significant load imbalance between concurrent entities, then consider us-ing the static round-robin shuffle to obtain easily accessible load balancing.

---

**Algorithm 4** Static Round-Robin Shuffle

1: **Input**
2:     $row\_old$                                                   ▷ sorted row buffer
3:     $segmentNum$                    ▷ The number of segments to shuffle
4: **Output**
5:     $row\_new$                               ▷ new row buffer after shuffle
6:     new2old                          ▷ the map of new id to old id
7: **procedure** SRRS($old\_array, segmentNum$)
8:     $num \leftarrow vertexNums/segmentNum$
9:     **for** $tid = 0; tid < segmentNum; tid + +$ **do**
10:        **for** $j = 0; j < num; j = j + 1$ **do**
11:            $row\_new[tid * num + j] \leftarrow row\_old[tid + segmentNum * j]$
12:            $new2old[tid * num + j] \leftarrow tid + segmentNum * j$
13:        **end for**
14:    **end for**
15:    **return** $row\_new, new2old$
16: **end procedure**

---

## 2.5 Block Search Bottom Up

In the Bottom-up algorithm, each iteration is scanned sequentially through the visit bitmap to find unsivited vertices. After several Bottom-up iterations, the number of unsivited vertices will be drastically reduced and sparsely distributed. Sequential scanning of the visit bitmap is inefficient. We propose a block search based Bottom UP algorithm [15], as shown in Algorithm 5,where 64 vertices form a block and are loaded into a general register so that the binary processing algorithm can quickly find the unsivited vertices in the register. The method also skips access to already traversed vertices at block granularity, lines 27 of the algorithm. In addition, we find that we can compress the three bitmaps used by the Bottom UP algorithm into only two, lines 23-48 of the algorithm, and the entire algorithm kernel is optimized for register processing and read operations on cache, with the write operations on cache reduced to one, occurring after the overall processing is completed on a block-by-block basis, lines 25-46 of the algorithm. In addition, because processing in blocks increases the proportion of effective computations, we merged two separate sections of degree-aware code into one, lines 31-44 of the algorithm. The above optimizations reduce access to cache, improve branching efficiency, and significantly improve the efficiency of single-core computation.

In summary, the above optimizations have been performed from the perspectives of reducing redundant memory access, improving NUMA memory access locality, improving multi-core load balancing, and improving single-core caching and branching efficiency, but have not yet touched an important cause of the severe random memory access of graph applications, the high random memory access due to low degree vertices. We transformed some random memory access into sequential memory access by edge tree optimization, which significantly reduced the impact of random memory access on performance.

---

**Algorithm 5** Block Search Bottom-up optimization

1: **Input**
2:     $G = \{V, E\} = \{V, (A^{IN+}, A^{IN-})\}$
3:     vertex_start
4: **Output**
5:     parent[N]                          ▷ the parent info of every vertex
6:
7: **procedure** BLOCKSEARCHUNVISITEDVERTEX($block\_no\_visit, pos$)
8:     $uint64\_t \; bit\_no\_visit \leftarrow block\_no\_visit \& (-block\_no\_visit)$
9:     $uint64\_t \; mask \leftarrow bit\_no\_visit - 1$
10:    $pos \leftarrow \_\_builtin\_popcount(mask)$
11:    $block\_no\_visit \leftarrow block\_no\_visit \& \sim bit\_no\_visit$      ▷ set the lowest unvisited vertex to visited
12:    **return** $pos, block\_no\_visit$
13: **end procedure**
14:
15: **procedure** BSB-BFS($G, vertex\_start$)
16:    $visit[v] \leftarrow 0, \forall v \in V$      ▷ the visit bitmap, 1 means the vertex has been traversed,0 means not
17:    $visit[vertex\_start] \leftarrow 1$
18:    $parent[v] \leftarrow -1, \forall v \in V$
19:    $parent[vertex\_start] \leftarrow vertex\_start$
20:    $flag \leftarrow 1 \rhd 1$ means the traversal has not been completed, 0 means not
21:    $BlockNum \leftarrow vertex\_num/BlockSize$      ▷ Total vertices divided by block width.
22:    **while** $flag == 1$ **do**
23:        $flag \leftarrow 0$
24:        **for** $i \leftarrow 0$ to $BlockNum$ **in parallel do**      ▷ The current layer is traversed, taking each block in turn.
25:            $block\_visit \leftarrow getBlock(visit, i) \rhd$ Get the access status of the ith block and store it in the block_visit register.
26:            $block\_no\_visit \leftarrow \sim block\_visit$      ▷ The current block has no traversed and undetected vertices.
27:            **while** $block\_no\_visit \neq 0$ **do**      ▷ The current block has no untraversed and undetected vertices.
28:                $pos \leftarrow$ BLOCKSEARCHUNVISITEDVERTEX($block\_no\_visit$)
29:                $w \leftarrow i * BlockSize + pos$
30:                $v \leftarrow A^{IN+}(w)$
31:                **if** getBit(visit,v) **then**
32:                    $block\_visit \leftarrow block\_visit | (1 << pos)$
33:                    $parent[w] \leftarrow v$
34:                    $flag \leftarrow 1$
35:                **else**
36:                    **for** $v \in A^{IN-}(w)$ **do**
37:                        **if** getBit(visit,v) **then**
38:                            $block\_visit \leftarrow block\_visit | (1 << pos)$
39:                            $parent[w] \leftarrow v$
40:                            $flag \leftarrow 1$
41:                            **break**
42:                        **end if**
43:                    **end for**
44:                **end if**
45:            **end while**
46:            $writeBlock(visit\_new, i, block\_visit)$      ▷ Write the updated visit state of the iTh block back to visit_new bitmap.
47:        **end for**
48:        swap(visit, visit_new)
49:    **end while**
50:    **return** $parent[N]$
51: **end procedure**

---

# 3 DATA RELEVANCE OF THE ALGORITHM

Graph algorithms (BFS, PageRank, etc.) and traditional high-performance numerical algorithms (matrix multiplication, FFT, convolution, etc.) have completely different architectural features, but there is no work yet to explain why this difference arises. The work on parallel tuning is prone to some optimization pitfalls. In the following, we propose a new classification of algorithms to explain the difference between these two classes, which is used later to illustrate the effectiveness of edge-tree graph traversal proposed in this paper.

## 3.1 Data Independent Algorithm(DIA)

**Definition**: The runtime memory access behavior of an algorithm does not depend on the specific value of any memory cell.

Memory ordering of data-independent algorithms is determined at compile time. The memory ordering does not change regardless of the data values stored in the memory cell. This good property leads to the fact that such algorithms can be easily accelerated by the compiler or by manually adjusting the order of memory accesses to improve the regularity and locality of the accesses, and can be easily accelerated using hardware. Traditional high-performance numerical algorithms (matrix multiplication, FFT, convolution) fall into this category, which are well established. As shown in Algorithm 6, the common optimization methods use loop unroll, loop exchange, tile, SIMD, prefetching, and systolic array, etc.

---

**Algorithm 6** Matrix Multiplication

1: **for** $i = 0; i < N; i + +$ **do**
2:    **for** $j = 0; j < N; j + +$ **do**
3:       **for** $k = 0; k < N; k + +$ **do**
4:          $c[i][j] \leftarrow c[i][j] + a[i][k] * b[k][j]$
5:       **end for**
6:    **end for**
7: **end for**

---

### 3.2 Data Correlation Algorithm(DCA)

**Definition:** The runtime memory access behavior of an algorithm depends on the specific value of a certain storage unit.

The intrinsic feature of data correlation algorithms is that the runtime state depends on some stored value, which leads to the fact that optimization methods for data-independent algorithms are generally ineffective for data correlation algorithms. Optimization of data-correlation algorithms is more difficult than optimization of data-independent algorithms. A typical representative of this class of algorithms is the graph algorithm (BFS, SSSP, PageRank, etc.). In addition to graph algorithms falling into this category, a large number of applications in data centers also fall into this category, and data centers are generally more focused on high throughput, hence this paper uses the term high throughput computing(HTC) [16] as a counterpart to high performance computing(HPC). There are two points to note, 1) Although a large number of data correlation algorithms have random access to memory features, data correlation algorithms do not always have random access features to memory. Depending on the contents of the storage unit, data correlation algorithms can exhibit both sequential and random memoy access features. The first for loop in Algorithm 7, if the value of childId in edgelist is continuously increasing, then the parent is sequentially accessed, and vice versa. This is a very important difference between data correlation and data-independent algorithm. This means that if an algorithm is of the data correlation type, the same piece of code does not need to be changed at all, and the memory order in which the algorithm is accessed can be changed simply by changing the arrangement of the data, leading to different performance results. 2) Random access data correlation algorithms are not necessarily cache-unfriendly. If the range of random accesses is smaller than the capacity of the cache, then these random accesses can also be hit in the cache. Such random accesses still have good cache locality. The second for loop in Algorithm 7, although the access to the bitmap is randomly accessed, still has good cache locality because the bitmap can be placed in the cache entirely.

Therefore, the cache friendliness of the data correlation algorithm is used here to further subdivide the algorithm, defined as cache-friendly and cache-unfriendly data correlation algorithms, respectively.

### Cache Friendly Data Correlation Algorithm(CFDCA)

Cache-friendly data correlation algorithms still have good memory access locality and can be further divided into sequential access cache-friendly data correlation algorithms and random access cache-friendly data correlation algorithms. The nature of cache-friendly data correlation algorithms approximates data-independent algorithms, and the same optimization methods for data-independent algorithms generally apply to cache-friendly data correlation algorithms. However, compilers do not perform compiler-level automatic optimizations like data-independent algorithms, because current compilers can only perform conservative optimizations and are not able to recognize runtime memory access locality. For cache-friendly data correlation algorithms, programmers are generally required to manually specify compiler optimization strategies to improve performance.

### Cache Unfriendly Data Correlation Algorithm(CUDCA)

Cache-unfriendly data-correlation algorithms exhibit truly random accesses, with very fine granularity, such that the range of addresses for two adjacent accesses exceeds the capacity of the cache. The characteristic makes memory access behavior difficult to predict at compile and run time. Graph processing algorithms fall strictly into this category. Optimization methods for data-independent algorithms are generally based on the regularity of memory access, and the locality of memory access can be improved by simply changing the control flow, but these methods do not change the true random memory access properties of data correlation algorithms after they are used to cache unfriendly memory access locality, cache-unfriendly data correlation algorithms generally rely heavily on data preprocessing. Most of the optimization methods used in the related work need to be implemented with the corresponding data preprocessing.

In summary, A detailed comparison of the above algorithms is shown in TABLE 1. From the above analysis, we can establish a clearer framework for algorithm optimization, avoid some optimization pitfalls. At the same time, we can also see that cache-friendly data correlation algorithm is a special class of data correlation algorithm, which is intrinsically data correlation algorithm, but are similar in nature to data-independent algorithm, and data-independent algorithm optimization tools can generally be used directly for cache-friendly data correlation algorithm. Cache-friendly data correlation algorithm appear to bridge the gap between cache-unfriendly data correlation algorithm and data-independent algorithm. This inspires us that if we are able to transform cache-unfriendly data correlation algorithm into cache-friendly data correlation algorithm, then we can make use of our familiar optimization methods and experience related to data-independent algorithm.

---

**Algorithm 7** Taverse EdgeList Only Leaf Vertex(TEOLV)

1:  **Input**
2:      $struct < src, dst > edgelist[M]$
3:  **Output**
4:      $parent[N]$                                              ▷ the parent info of every vertex
5:
6:  **procedure** TEOLV($edgelist, parent$)
7:      /*
8:      **for** $i = 0; i < M; i++$ **do**     ▷ CFDCA if dst is sequentially increasing, otherwise maybe CUDCA
9:          $parentId \leftarrow edgelist[i].src$                        ▷ Sequential memory access
10:         $childId \leftarrow edgelist[i].dst$
11:         $parent[childId] \leftarrow parentId$
12:     **end for**
13:     */
14:
15:     **for** $i = 0; i < M; i++$ **do**     ▷ CFDCA if dst is sequentially increasing and the bitmap can be filled into the cache, otherwise maybe CUDCA
16:         $parentId \leftarrow edgelist[i].src$
17:         $childId \leftarrow edgelist[i].dst$
18:         **if** (bitmap(parentId)) **then**            ▷ random memory access but cache friendly
19:             $parent[childId] \leftarrow parentId$
20:         **end if**
21:     **end for**
22:     **return** $parent[N]$
23: **end procedure**

TABLE 1 The Comparison of Algorithm Type

|  | DIA | CFDCA | CUDCA |
|---|---|---|---|
| Data Dependency | Explicitly Independent | Implicitly Independent | Runtime Dependency |
| Cache Locality | High | High | Low |
| Random Access | Low | Low | High |
| Prefetch | Easy | Easy | Hard |
| Load Balancing | Easy | Easy | Hard |
| Compiler Optimization | Auto | Set Manually | Can't |
| CPU Friendly | Yes | Yes | No |
| Tuning Difficulty | Easy | Easy | Hard |
| Typical Application | GEMM/FFT | | BFS/PageRank |
| Computing Category | HPC | HTC/HPC | HTC |

## 4 EDGE TREE VIEW OF THE GRAPH

### 4.1 Small Degree Vertices and Random Access

The optimization of data correlation algorithms is closely related to the properties of the data. A change in the data layout will change the memory access behavior of the algorithm. In many cases, a change in the data layout will result in a larger performance improvement than the optimization of the algorithm itself. Some related work has exploited the relevant properties of power-law graphs with large degree vertices[13],[14], but no related work has investigated the small degree vertices. We find that small degree vertices are also one of important cause of the high random access to the graph processing. Nature graph conforms to skewed power-law degree distribution[5],[17]. Most vertices have relatively few neighbors while a few have many neighbors. The kronecker graph [18], a common power graph generator, is heavily used in the graph research field and the Graph500 uses it as an input graph as well. As shown in TABLE 2, the Kroneker graph (SCALE=26) has 63% of small vertices, with 51.1% of isolated vertices(VZs) and 12.4% of vertices with degree 1 (TLs). In the graph processing algorithm, if the neighbor vertices of a vertex contain small degree vertices, then obviously the memory access of two neighbor vertices adjacent to each other is probabilistically true random access. And this random memory access is fine-grained in terms of vertices. This problem will become more worse under the multi-core, NUMA architecture of the existing architecture, and the existing architecture faces severe challenges.

### 4.2 Edge Tree View of Graph

Our further analysis shows that the graph data has not only a large number of small degree vertices, but also a large number of low degree vertices. Several low degree vertices form a tree, and the graph data has a large number of such trees, which we define as edge tree(ET). As shown in Fig. 1, the left and right graphs are the same graph. The graph on the right simply adjusts the position of the lower vertices, the graph on the left looks very chaotic, and the graph on the right has structure. We define this kind of graph on the right as an edge tree view of the graph(ETVG). The original graph can be understood as consisting of core subgraphs and a large number of edge trees.

After defining the concept of edge tree, we mark the vertices in the graph, which can be divided into five categories:

● Core Internal Vertex (CI)
   Such vertices are located in the core graphs and are not connected to any edge trees. Such vertices are shown in white on the diagram.

● Core Edge Vertex (CE)
   Such vertices are located in the core graphs and are connected with some edge trees. Such vertices are shown in red in the diagram.

● Tree Internal Vertex (TI)
   Non-leaf vertices on the edge trees. Such vertices are shown in green in the diagram.

● Tree Leaf Vertex(TL)
   The leaf vertices on the edge trees. That is, vertices of degree 1 in the original graph. Such vertices are indicated in blue in the graph.

● Vertex Zero (VZ)
   Isolated vertices in the original graph. Such vertices are shown in black in the diagram

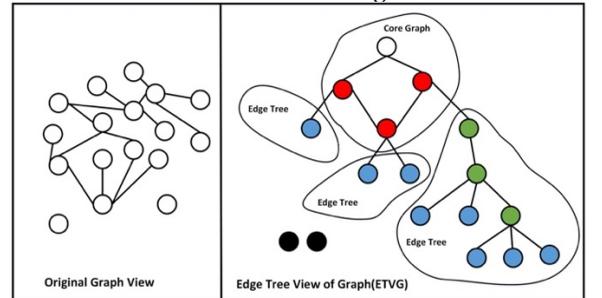

Fig. 1 Graph and Its Edge Tree View

### 4.3 Edge Tree Vertex Classification Algorithm

The goal of the edge tree vertex classification algorithm is to classify the vertices in the original graph, labeled as CORE_INTERNAL, CORE_EDGE, TREE_INTERNEL, TREE_LEAF, and VERTEX_ZERO, respectively. As shown in Algorithm 8, initially all vertices are of type CORE_INTERNAL by default, and then they are marked from the bottom up from the leaf vertices. Vertices of degree 1 and 0 are marked with TREE_LEAF and VERTEX_ZERO, respectively, and then the TREE_LEAF vertices and their neighbors are deleted from the graph. Then select a vertex of degree 1 or 0, mark it as a vertex of type TREE_INTERNEL. Then delete the vertices of type TREE_INTERNEL and their neighbor edges from the graph, and repeat the process until there are no vertices of type TREE_INTERNEL, and complete the marking of the vertex of type TREE_INTERNEL. Finally, by comparison with the original graph, if the degree of the vertex has changed, then the vertex is marked as a vertex of type CORE_EDGE. The remaining vertices are vertices of type CORE_INTERNAL, which are set initially and do not need to be processed. By controlling the height of the edge tree, different MH divisions are obtained. MH = 0 division, only vertices of type TREE_LEAF and VERTEX_ZERO are marked. There are no vertices of type TREE_INTERNAL in the edge tree view. The core graph in the edge tree view is directly connected to a large number of leaf vertices. We call this special case as edge leaf view of graph (ELVG). The algorithmic complexity of the preprocessing at this point is O(V).

---

**Algorithm 8 Edge Tree Vertex Classification Algorithm**

```
1: Input
2:   G ⇒ {V,E}
3:   height▷ MH,Maximum height of the edge tree to be classified, default 0 classify leaf
     nodes only;-1 indicates a fully preserved classified edge tree of the original graph
4: Output
5:   vertex_type[N]                              ▷ array of storing the type of every vertex
6:
7: procedure ETVCA(G)
8:   vertex_type[N].N init to CORE_INTERNAL
9:   memcpy(vertex_num_pre, G.vertex_num, N)     ▷ copy array of storing the number of
     neighbors
10:  memcpy(vertex_num_new, G.vertex_num, N)     ▷ copy array of storing the number of
     neighbors
11:  flag ← 0    ▷ 0 indicates that there are not vertices in the current height in the edge
     tree
12:
13:  for v ∈ V do                                ▷ classify TYPE.ZERO and TREE.LEAF
14:    if vertex_num_pre[v] == 0 then
15:      vertex_type[v] ← TYPE.ZERO
16:    else if vertex_num_pre[v] == 1 then
17:      vertex_type[v] ← TREE.LEAF
18:      for each w adjacent to v do
19:        vertex_num_new[w] ← vertex_num_new[w] − 1
20:      end for
21:      flag ← 1
22:    end if
23:  end for
24:
25:  height_current ← 1
26:  while flag == 1 ∧ (height == −1 ∨ height_current <= height) do    ▷ classify
     TREE.INTERNAL
27:    flag ← 0    ▷ start processing vertices in edge trees of height_current
28:    memcpy(vertex_num_pre,vertex_num_new,N)
29:    for v ∈ V do
30:      if (vertex_type[v] == CORE_INTERNAL) ∧ (vertex_num_pre[v] == 0 ∨
         vertex_num_pre[v] == 1) then
31:        vertex_type[v] ← TREE.INTERNAL
32:        for each w adjacent to v do
33:          vertex_num_new[w] ← vertex_num_new[w] − 1
34:        end for
35:        flag ← 1
36:      end if
37:    end for
38:    height_current ← height_current + 1
39:  end while
40:
41:  for v ∈ V do                                ▷ classify CORE.EDGE
42:    if (vertex_type[v] == CORE_INTERNAL) ∧ (vertex_num_new[v] ≠
       G.vertex_num[v]) then
43:      vertex_type[v] ← CORE.EDGE
44:    end if
45:  end for
46:
47:  return vertex_type[N]
48: end procedure
```

---

### 4.4 Properties of The Edge Tree View

1. The two types of vertices, Core Internal Vertex and Core Edge Vertex, make up the core graph, which is a smaller subgraph of the same nature as the original graph and still has true random access properties.

2. Tree Internal Vertex and Tree Leaf Vertex make up the edge trees. These types of vertices, although contributing heavily to random access in the original graph, have the potential to be optimized for sequential memory access because the tree is a special data structure.

3. At most one Core Edge Vertex is connected to the root vertex of each edge tree. As shown in Fig. 1. each edge tree corresponds to a unique Core Edge Vertex (red vertex).

4. Vertices in each edge tree are not connected to any

Core Internal Vertex type.

5. The edge tree has all the properties of a tree structure. The parent vertex of each vertex is unique.

### 4.5 The Height of Edge Tree

Given a graph G, if the fathers are sequentially partitioned into edge trees starting from the leaf vertices, the number of vertices in the core graph gradually decreases and the number of vertices in the edge tree gradually increases, and eventually the two sets converge to some fixed value. Define the maximum height of all edge trees at this point as the Peak Height(PH) of the edge tree of the graph. Note that the PH of the kronecker graph is very small, e.g. a kronecker graph with scale=26 has a PH of 2. For a given Max Height(MH, less than or equal to PH), divide the vertices in the graph as far as possible onto the edge tree, but ensure that the maximum height of all edge trees does not exceed MH, called the MH edge tree division of the graph. MH = 0 is a special case where only leaf vertices are divided onto the edge tree. For any graph, get the relevant parameters in its edge tree view and many optimization issues will become clear. For example, Fig. 2 are the all MH divisions of Fig. 1. TABLE 2 is the Kronecker graph with scale=26, edgefactor=16, and the number of vertices of each type under different MH divisions. We can see that the proportion of TL and VZ is very high, accounting for 63%, TI type accounting for 0.03%. Only 37% of the vertices in the core graph (CI and CE). TABLE 3, MH=2, scale=26 graph contains a total of 2484171 Core Edge Vertex connected edge trees, these edge trees contain a total of 8332878 vertices, on average each edge tree contains 3 vertices, the largest edge tree contains 9685 vertices, the smallest edge tree contains 1 vertex. This indicates that the edge tree is severely sparse and is an important cause of random access to the memory. This inspires the possibility of special treatment of these low-degree vertices in edge trees individually to improve performance.

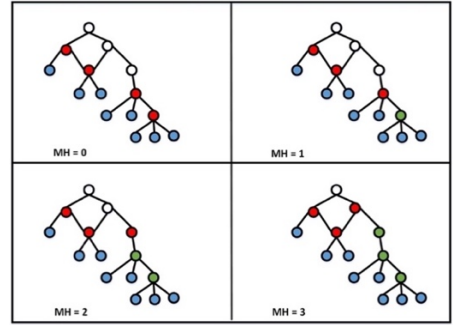

Fig. 2 The Divisions under Different MHs

TABLE 2 The Number of Vertices in Each Type under Different MH classifications for Scale=26,Edgefactor=16

| MH | CI | CE | TI | TL | VZ | Total |
|----|----|----|----|----|----|-------|
| 0 | 21970533 | 2502108 | 0 | 8332198 | 34304025 | $2^{26}$ |
| 1 | 21966322 | 2484225 | 22094 | 8332198 | 34304025 | $2^{26}$ |
| 2 | 21966312 | 2484171 | 22158 | 8332198 | 34304025 | $2^{26}$ |

TABLE 3 The Distribution of the Number of Vertices Contained in All Edge Trees Connected by Core Edge Vertex in the Kronecker Graph with Scale=26,Edgefactor=16, MH=2

| Edge Tree Number | Ave | Max | Min | Total |
|------------------|-----|-----|-----|-------|
| 2484171 | 3 | 9685 | 1 | 8332878 |

## 5 EDGE TREE TRAVERSAL ALGORITHM

The original graph in the edge tree view is partitioned to consist of some core subgraphs and edge trees. Considering that the father of the leaf vertex can be determined before the algorithm run, the father information of the leaf vertex can be made into a lookup table in the data pre-processing stage, so that the algorithm only needs to process the core graphs and not the leaf vertices. Thus it improves the performance of the BFS algorithm. However, there are two key problems with this approach: 1) The leaf vertices are not re-visited during the run of the algorithm, losing the generality of the BFS algorithm as a basic pattern for graph processing algorithms. Some graph processing algorithms need to update the state of all vertices in each iteration, such as PageRank, etc. 2) The performance improvement brought by the above approach may come from this part of the removed access to the memory and that not an optimization of random memory access. Is there a method that simultaneously processes leaf vertices during graph algorithm traversal that guarantees both generality and high performance? We propose an edge tree breadth-first traversal method to solve this problem, which is a method that guarantees both generality of the BFS algorithm model and high performance.

### 5.1 Data Structure and Layout

#### 5.1.1 Master Data Structure

The data structure of graph is stored in the well-known Compressed Sparse Row (CSR) format, adopted by most graph algorithms and graph processing systems. The CSR consists of two lists as shown in Fig. 3. The adjacency list stores the neighbor information and its size is bounded by the number of edges. In the row list, it stores the first neighbor's pointer of each vertex. The CSR format allows streaming access of all neighbors for each vertex. The main data structure still uses the CSR. The vertices in the edge tree view can be divided into two categories, one is the vertices in the core subgraph, which is further subdivided into two types CORE_INTERNEL and CORE_EDGE. The other category is the vertices in the edge tree, further subdivided into the types TREE_INTERNEL, TREE_LEAF, and VERTEX_ZERO. The data layout in the CSR is as follows.

- Row array. Place the vertices in the core subgraph to the left of Row and the vertices in the edge tree and isolated vertices to the right of Row. As shown on the right side of Fig. 3, a, d, and e are the vertices in the core subgraph placed to the left, and b, f, and c are the vertices in the edge tree placed to the right.
- Col array. All neighbor vertices of each vertex are also separated by type, with vertices in the core subgraph placed on the left and vertices in the edge tree on the right. As shown on the right side of Fig. 3, d and e of the neighbors of vertex a are the vertices in the core subgraph to the left and b and f are the vertices in the edge tree to the right.

The above proposed is a layout idea, the BFS algorithm comes without further adjustment of the layout. Other algorithms can further adjust the layout according to the characteristics of the algorithm. This data structure and layout has the following advantages.

- Guaranteed generality and compatibility of data structures and graph algorithms. The data structure is still in the CSR format, just adjusting the layout of the data in the CSR, and the other graph algorithms and optimizations work with little to no change. Restoring the layout is also easy.
- The storage of vertices in all edge trees is continuously incremental and can be processed sequentially using the CFDCA algorithm.
  Previous CSR data layout in which vertices in the edge trees and vertices in the core graphs are stored together in a mixture, cannot process the vertices in the edge tree sequentially. In our proposed layout, all the vertices in the edge tree are on the right side of the CSR and their numbering is continuous, enabling sequential processing using the CFDCA algorithm to improve performance.
- There is little impact on the performance of different optimization methods. For example, the degree-aware optimization mentioned in the related work requires that the high-degree vertices in each neighborhood are on the left and the low-degree vertices are on the right, and the layout here also satisfies this condition, with the left core being the high-degree vertices in the core graph and the right side being the low-degree vertices in the edge tree.
- Low data pre-processing complexity and cost
  Although graph applications do not generally require performance for data preprocessing, the data preprocessing of the edge tree algorithm proposed in this paper still has a low algorithmic complexity. The edge tree vertex classification algorithm has a low algorithmic complexity. The algorithmic complexity of adjusting the CSR layout is O(V+E) when the type of vertices is obtained. Moreover, these preprocessing algorithms are cache-friendly data correlation algorithms, easy to parallelize and easy to optimize.

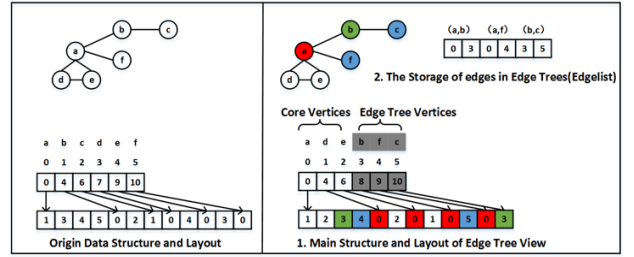

Fig. 3 The Data Structure and Layout of Edge Tree View

#### 5.1.2 Storage and Representation of Edge Trees

The main data structure and layout are designed with the idea of ensuring generality and compatibility. The edgelist data structure is used to store the edge tree. The edge trees are stored using the edgelist data structure. The edgelist represents the edges in the edge trees as arrays of elements (src,dst). The src and dst represent the starting vertex and end vertex, respectively, and their values range from the number of vertices in the CSR data structure, and since all vertices in the edge trees are to the right of the row, the numbering of all vertices in the edge tree is ordered incremental. This means that different algorithms can maintain sequential incrementation of src or dst at adjacent positions in edgelist by a simple layout according to the memory access characteristics of their algorithms. This is a very important property. By taking advantage of this sequential nature, the DCA algorithm can be optimized from CUDCA to CFDCA, and by reading the edgelist sequentially, all edges in the edge trees can be processed sequentially to improve the memory access efficiency and performance. For the BFS algorithm, e.g. Fig. 3 top right corner, is the storage for the edge tree.

## 5.2 Edge Tree BFS Algorithm(ET-BFS)

As shown in Algorithm 9, if the start vertex of the traversal is on the edge tree, the vertex is of type TREE_INTERNAL or TREE_LEAF, then it is necessary to first traverse the edge tree alone and return the corresponding CORE_EDGE vertex of the edge tree (this vertex is unique by property 3). For MH=0 division, only leaf vertices are marked, so this step can be omitted. Next, the core subgraphs and the edge trees are processed separately. The BFS_CORE indicates any BFS algorithm, except that the size of the processing graph data is changed from the original graph to the smaller core subgraph, and the vertices in the edge trees will no longer participate in the processing. Then, the processing of vertices in the edge trees can be completed by simply traversing the edge tree edgelist through sequential memory access. The core idea of the method is to treat vertices in the edge tree (vertices of low degree) and vertices in the core subgraphs differently, using the CSR data structure to process vertices in the core subgraph and using the edgelist data structure to process vertices in the edge tree. The previous approach did not deeply recognize the different nature of the vertices in the graph and treated all vertices uniformly, resulting in low-degree vertices bringing a large number of random memory access and not taking advantage of the potential for sequential memory access that exist in the low-degree vertices.

---

**Algorithm 9** Edge Tree BFS Algorithm

1: **Input**
2:     G = {V,E} = {Core,Tree}
3:     $struct < src, dst, ... > edgelist[M]$
4:     vertex_start            ▷ the starting vertex of the traversal
5: **Output**
6:     parent[N]            ▷ the parent info of every vertex
7:
8: **procedure** ET-BFS($G, edgelist, vertex\_start$)
9:     **if** $vertex\_start \in Tree$ **then**
10:        $(vertex\_start\_new, parent) \leftarrow traverse\_return\_core\_edge(G, vertex\_start)$
       ▷ traverse the edge tree where the start vertex located and return the corresponding CORE_EDGE vertex of the edge tree.
11:        $vertex\_start \leftarrow vertex\_start\_new$    ▷ update the vertex start
12:     **end if**
13:
14:     BFS_CORE(G,vertex_start)     ▷ CUDCA algorithm. Just process core graph
15:     TRAVERSE_EDGELIST(edgelist)    ▷ CFDCA algorithm. Just process edge trees
16:     **return** $parent[N]$
17: **end procedure**

---

## 5.3 Edge Tree Processing Algorithm

As shown in Algorithm 10, the processing of edges in all edge trees is achieved by traversing the edgelist. Some edge trees will not be traversed during the traversal process because they are not located on the connected subgraph where the start vertex is located and belong to another connected subgraph. However, the graph500 requires that the vertex with a non-empty parent value must be on the traversal spanning tree of the starting vertex as the root vertex, so it need to determine whether the edges in edgelist are on the traversal spanning tree edge of the starting vertex as the root vertex, only the vertices that can be traversed need to update their father information. The core_edge indicates the Core Edge Vertex of the edge tree where (src, dst) is located. If the core_edge has been visited, then it means dst can be traversed and its father can be updated. A caveat here is that the dst in edgelist needs to be arranged in ascending order and the bitmap can be filled into the cache so that the algorithm can become a CFDCA type algorithm. The cache is localized well and easy to optimize. Our proposed CSR storage layout naturally makes dst ascending because the numbering of all vertices in the edge trees of a CSR data structure is incremented serially, and the vertex numbering in edgelist is the numbering of the CSR, so the dst values of the neighboring positions in edgelist are also incremented serially. This is a very important point, because if the dst of adjacent positions in the edgelist is not continuously incremented, then the accesses to the parent array will not be continuous, and Algorithm 10 will degenerate into a CUDCA type algorithm, still with a large number of random accesses, which experiments show will not lead to performance improvement. For the edge leaf view, MH=0, and only leaf vertices are extracted, then the algorithm can be judged directly using src instead of core_edge, as shown in Algorithm 7 TEOLV, which has a simpler form. According to the power-law rate nature of the natural graph, a large number of vertices in the edge trees are leaf vertices, such as TABLE 2 and TABLE 3. TE-OLV algorithm form is not only simple, but also experimental results show that it has high performance at the same time.

---

**Algorithm 10** Taverse EdgeList Edge Tree(TEET)

1: **Input**
2:     $struct < src, dst, core\_edge > edgelist[M]$
3: **Output**
4:     parent[N]            ▷ the parent info of every vertex
5:
6: **procedure** TEET($edgelist, parent$)
7:     **for** i = 0; i¡ M; i++ **do**       ▷ CFDCA algorithm when the childIds are in ascending order and the bitmap can be filled into the cache(high performance), otherwise CUDCA(low performance)
8:        $parentId \leftarrow edgelist[i].src$
9:        $childId \leftarrow edgelist[i].dst$
10:        $core\_edge \leftarrow edgelist[i].core\_edge$
11:        **if** bitmap(core_edge) **then**     ▷ random memory access but cache friendly
12:           $parent[childId] \leftarrow parentId$    ▷ sequential memory access,if the childIds are in ascending order
13:        **end if**
14:     **end for**
15:     **return** $parent[N]$
16: **end procedure**

# 6 EXPERIMENTAL EVALUATION

## 6.1 Experiment Setup

In order to evaluate the validity of the methodology, we used computing platforms provided by the three leading vendors of supercomputing and data center, Intel, AMD, and ARM. As shown in TABLE. 4, for their specific configurations. The Intel E5-2683 is mainly used to evaluate the effectiveness of the method, the AMD EPYC 7452 is the latest processor for evaluating the maximum performance, and the ARM processor is used to evaluate the potential of the ARM architecture as an emerging server architecture. Unless otherwise noted, both Intel and AMD are compiled using the icc 19.0.0 compiler and ARM processors are compiled using gcc 8.3.0. The test dataset was generated using kronecker graph generator from the Graph500 benchmark with the parameters set to default values (A = 0.57, B = C = 0.19, D = 0.05). The kronecker graph generator can be adjusted by entering parameters such as scale and edgefactor, where the scale parameter indicate the scale of the vertices of the graph, and the edgefactor indicates the average degree of each vertex, where the default value of Graph500 is 16. The generated graph data satisfies the power-law distribution and contains the number of vertices in $2^{scale}$ and the total number of edges in $2^{scale}*edgefactor$. According to Graph500, the performance is represented by giga-traversed edges per seconds (GTEPS). 64 source vertices are randomly selected to execute the BFS algorithm, and then the average of the results from these 64 vertices is taken as the final performance.

| Platform | Intel | AMD | ARM |
|---|---|---|---|
| Operation System | CentOS7 | CentOS7 | CentOS7 |
| CPU | Xeon E5-2683 v3 | EPYC 7452 | HTC Centriq 2434 |
| CPU speed | 2.00 GHz | 2.35 GHz | 2.30 GHz |
| Socket(s) | 2 | 2 | 1 |
| Cores per socket | 14 | 32 | 40 |
| L3 cache | 35 MB | 128 MB | 50 MB |
| Memory size | 384 GB | 256 GB | 384 GB |
| Memory type | DDR4 | DDR4 | DDR4 |
| TDP | 120 W | 155 W | 110 W |

TABLE 4 Configure information

## 6.2 Performance of Different MH Divisions

The machine used for the experiments in this section is the E5-2683, and this section uses all the optimization methods that is Hybrid+RmZero+RoundRobin+NumaAware+DegreeAware+BlockSearch+ET-BFS. The graph is set to scale=26, edgefactor=16 to study different MH divisions' performance. As shown in Section 4.5, the higher the MH, the more vertices will be partitioned to the edge tree. The PH of the graph for SCALE=26 is 2. MH=0 is a special case and can also be used to process leaves using Algorithm 7 TEOLV, which is also used here as a comparison. The Fig. 4 shows that the performance is almost identical under different MH divisions. This is due to the small percentage of TI vertices, which are basically leaf vertices. Considering the simplicity of the TEOLV algorithm, TEOLV is chosen as the object of study in the latter part of the paper.

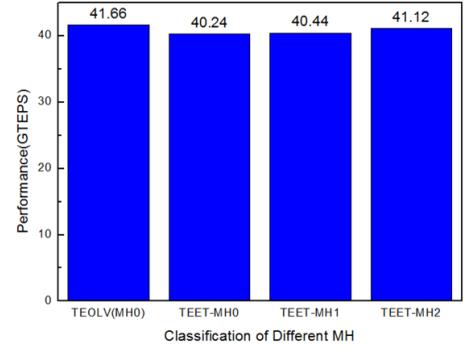

Fig. 4 The Performance of Classification of Different MH

## 6.3 Strong Efficiency

The trigger conditions for the validity of data-correlation algorithm and data-independent algorithm are also different. The optimization methods of a data-independent algorithm generally leads to performance improvements on top of other optimization methods, but some optimization methods of a data-correlation algorithm can only show an optimization effect when paired with a particular optimization method. A data correlation algorithm is said to be strongly effective if the optimization method of the algorithm can further improve performance on the basis of most other optimization methods. This section evaluates the strong effectiveness of the edge tree algorithm by adding edge tree optimization to any optimization method. The machine used in this section is E5-2683. Kronecker graph with SCALE=26 and Edgefactor=16. As shown in Fig. 5, each element of the X-coordinate represents the added optimization relative to the previous one. The first column of each X-coordinate represents the previous performance. The second column represents the addition of the full edge tree optimization (BFS_CORE and TEOLV) to it, and the third column represents the addition of BFS_CORE relative to the first column. For example, the first column of Hybrid indicates the initial use of Hybrid optimization only, the second column indicates the use of Hybrid+BFS_CORE+TEOLV, and the third column indicates Hybrid+BFS_CORE. As for RoundRobin, the first column indicates the use of Hybrid+RmZero+RoundRobin optimization, and the second column indicates the use of Hybrid+RmZero+RoundRobin+BFS_CORE+TEOLV, and the third column represents Hybrid+RmZero+RoundRobin+BFS_CORE. This diagram contains a wealth of information to see that edge tree algorithm are strongly effective algorithms that can deliver performance improvements based on any optimization methods.

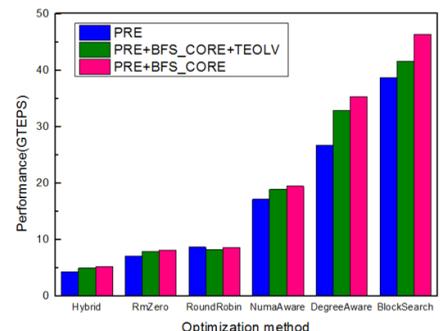

Fig. 5 The Strong Efficiency of ET-BFS Algorithm

## 6.4 Overall Performance Analysis

The machines used for the experiments in this section are E5-2683 , HTC Centriq and EPYC-7452, and the performance under different scales such as Fig. 6 was tested using kronecker graph with edgefactor = 16. where PRE indicates previous optimization, Hybrid+RmZero+ RoundRobin+NumaAware +DegreeAware+BlockSearch. The AMD EPYC 7452 performance values correspond to the right Y-axis, the rest of the platforms correspond to the left Y-axis.

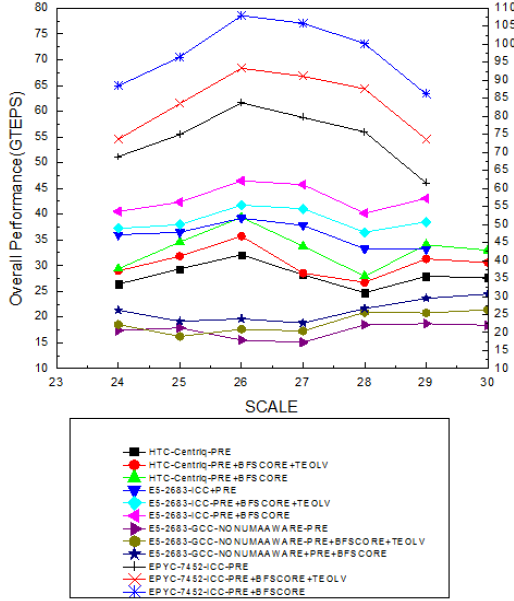

Fig. 6 Overall Performance

*Big Graph Data Efficiency*

Many optimization methods are effective for smaller scale graphs, but are ineffective for processing large graphs, and algorithms that are effective for large graphs are more difficult to design. A major reason for this is that as the size of the graph changes, the space required to represent the bitmap of the vertices also becomes larger, exceeding the capacity of the cache. Our proposed EdgeTree optimization algorithm still shows acceleration for large graphs because it decomposes the large graph into smaller core graphs. As in Fig. 6, on all platforms, the 28, 29, and 30 graphs show performance improvements relative to PRE.

*Performance Upper Bound*

The edge tree processing algorithm is of CFDCA type and easy to optimize. Different graph processing algorithms can tune the edge tree processing algorithm according to their own memory access characteristics. The core graph determines the upper bound on the performance of the edge tree processing algorithm optimization. For example, the complete edge tree algorithm BFS_CORE+TEOLV of Fig. 6 has an average performance gap of 4 GTEPS relative to BFS_CORE, which still has room for optimization. It is worth mentioning that although the full edge tree algorithm brings about an 8% performance improvement over the previous optimization, TEOLV is still an initial version of the code implementation that has not been fine-grained yet, just to illustrate the effectiveness of the edge tree algorithm with minimal implementation cost.

*Platform Performance Comparison*

The BFSCORE+TEOLV on the E5-2683, HTC-Centriq, and EPYC 7452 platforms improved all SCALE by an average of 8.0%, 8.7%, and 13.2%, respectively, relative to PRE. The BFSCORE on the E5-2683, HTC-Centriq, and EPYC 7452 platforms improved all SCALE by an average of 19.7%, 17.9%, and 31.8%, respectively. The EPYC 7452 platform has the strongest performance due to the use of the most advanced manufacturing process, huge capacity LLC, and the largest number of physical cores. The HTC-Centriq is essentially the same configuration as the E5-2583. Since the E5-2683 uses an ICC compiler by default and has two NUMA nodes, it has been optimized with NumaAware compared to the HTC-Centriq platform. We also tested with GCC and without NumaAware on E5-2683 for increased comparability. As shown in Fig. 6, HTC-Centriq Platform performance is on average 57.9% higher than the E5-2683, providing a significant performance advantage.

## 6.5 Scalability Analysis

This section tests the thread scalability of the edge tree BFS algorithm under different edgefactor with SCALE=26. The HTC Centriq platform is used to illustrate this scalability, considering that it has more cores. For example, when the Fig. 7 average degree is 16, high concurrent processing under 40 threads improves the performance by a factor of 24.43 over single threads, and the performance scales approximately linearly with increasing number of threads. The higher the number of edgefactors, the higher the performance. This is due to the nature of the Kronecker-generated graphs, which become less sparse as the average degree increases. The average performance of the algorithm can reach 56.23 GTEPS at an average degree of 32, which is better than 35.67 GTEPS at an average degree of 16 and 21.41 GTPES at an average degree of 8.

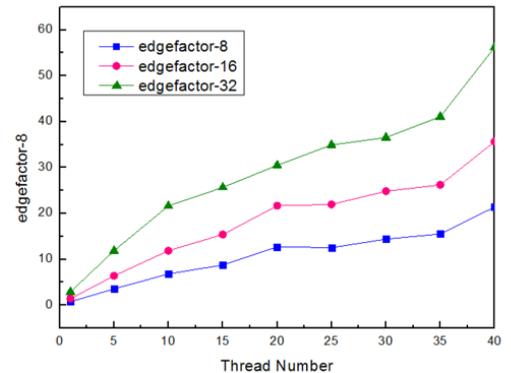

Fig. 7 The scalability of ET-BFS under Different Edgefactors

## 6.6 Cache Efficiency

The basic idea of the edge tree algorithm proposed in this paper is to decompose the random access data correlation algorithm into a smaller random access data correlation algorithm and a cache-friendly data correlation algorithm, which reduces the size of the random access data and improves the cache efficiency. This section uses Perf to obtain the LLC Cache Miss Rate in the multicore to observe this efficiency improvement. As in Fig. 8, the LLC cache miss rate of TEOLV is only half of that of BFSCODE, and the LLC cache miss rate of the complete ET-BFS algorithm is also reduced. This fully demonstrates that the optimization approach in this paper effectively improves the cache locality and the memory access efficiency.

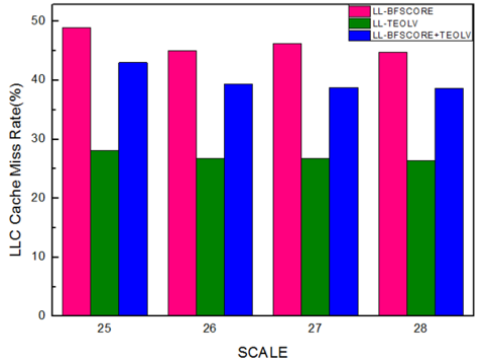

Fig. 8 LLC Cache Miss Rate

### 6.7 Performance and Power Consumption Comparison

As shown in TABLE 5, the performance and power consumption of the main platforms in the Green Graph500 ranking for this research area are listed for the same period. With the addition of the optimizations mentioned in this paper, the HTC Centriq platform has further improved its performance, ranking first on the 2019 Graph500 large dataset list. Compared to Tesla P100 GPU there is still a 1.59x performance power advantage. HTC Centriq, despite having only 1 numa node, still has 3.17 GTEPS higher performance than the previous representative work in the CPU space[13] (4-way machines) and even a 4.49x improvement in performance power consumption. Once again, the efficiency of the approach proposed in this paper is fully demonstrated.

TABLE 5 The Performance and Power Consumption Comparison of Different Platforms

| Reference[19] | Platform | Core | RAM(GB) | Scale | Edge factor | GTEPS | MTEPS | GreenGraph500 |
|---|---|---|---|---|---|---|---|---|
| This work | 1-way HTC Centriq 2434 | 40 | 384 | 30 | 16 | 34.49 | 282.70 | Nov 2019 #1 |
| Other | IBM Power8+ Tesla P100 | 66 | | 30 | 16 | 41.7 | 177.45 | Nov 2019 #2 |
| Other | IBM POWER8+ | 10 | | 30 | 16 | 13.2 | 66.0 | Nov. 2019 #6 |
| Other[13] | 4-way Xeon E5-4640 | 32 | 512 | 30 | 16 | 31.32 | 62.93 | Nov. 2019 #7 |
| This work | 2-way EPYC 7452 | 64 | 256 | 29 | 16 | 86.15 | | |

## 7 Conclusion

In this paper, we propose a new way of classifying algorithms into DIA, CFDCA, and CUDCA based on their runtime correlation and cache friendliness with data. The differences between data correlation algorithms and high performance numerical algorithms are expressed in depth, which can be useful for future data correlation algorithm optimization and architecture design. We find that small degree vertices are an important cause of high random memory access in graph processing, and propose a basic perspective of graph data understanding, which views graphs as consisting of core graphs and edge trees, and provides a basic analytical model for relevant research in the field of graph processing. Finally, we propose a general, easy-to-implement, and strongly effective breadth-first traversal algorithm for graph data, ET-BFS, which provides a new way of thinking for future optimization work in the field of supercomputing and graph processing. The experimental results show that it brings 19.7%, 17.9%, and 31.8% performance improvement on mainstream platforms such as E5-2683, HTC-Centriq, and EPYC 7452, respectively. The performance-power ratio on HTC-Centriq platform is 282.70 MTEPS/s, which is in the November 2019 Green Graph500 list ranked first in the world[19].

## Acknowledgment

## References

[1] A.-L. Barabási and R. Albert, "Emergence of Scaling in Random Networks," *Science*, vol. 286, no. 5439, pp. 509–512, Oct. 1999, doi: 10.1126/science.286.5439.509.

[2] T. H. Haveliwala, "Topic-sensitive PageRank: a context-sensitive ranking algorithm for Web search," *IEEE Trans. Knowl. Data Eng.*, vol. 15, no. 4, pp. 784–796, Jul. 2003, doi: 10.1109/TKDE.2003.1208999.

[3] A. Mislove, M. Marcon, K. P. Gummadi, P. Druschel, and B. Bhattacharjee, "Measurement and analysis of online social networks," in *Proceedings of the 7th ACM SIGCOMM conference on Internet measurement*, New York, NY, USA, Oct. 2007, pp. 29–42, doi: 10.1145/1298306.1298311.

[4] D. A. Bader and K. Madduri, "SNAP, Small-world Network Analysis and Partitioning: An open-source parallel graph framework for the exploration of large-scale networks," in *2008 IEEE International Symposium on Parallel and Distributed Processing*, Apr. 2008, pp. 1–12, doi: 10.1109/IPDPS.2008.4536261.

[5] J. E. Gonzalez, Y. Low, H. Gu, D. Bickson, and C. Guestrin, "PowerGraph: Distributed Graph-parallel Computation on Natural Graphs," in *Proceedings of the 10th USENIX Conference on Operating Systems Design and Implementation*, Berkeley, CA, USA, 2012, pp. 17–30, Accessed: Jan. 01, 2018. [Online]. Available: http://dl.acm.org/citation.cfm?id=2387880.2387883.

[6] J. Shun and G. E. Blelloch, "Ligra: A Lightweight Graph Processing Framework for Shared Memory," in *Proceedings of the 18th ACM SIGPLAN Symposium on Principles and Practice of Parallel Programming*, New York, NY, USA, 2013, pp. 135–146, doi: 10.1145/2442516.2442530.

[7] J. Shun, L. Dhulipala, and G. E. Blelloch, "Smaller and Faster: Parallel Processing of Compressed Graphs with Ligra+," in *2015 Data Compression Conference*, Apr. 2015, pp. 403–412, doi: 10.1109/DCC.2015.8.

[8] X. Zhu, W. Chen, W. Zheng, and X. Ma, "Gemini: A Computation-centric Distributed Graph Processing System," in *Proceedings of the 12th USENIX Conference on Operating Systems Design and Implementation*, Berkeley, CA, USA, 2016, pp. 301–316, Accessed: Mar. 01, 2018. [Online]. Available: http://dl.acm.org/citation.cfm?id=3026877.3026901.

[9] S. Grossman, H. Litz, and C. Kozyrakis, "Making pull-based graph processing performant," in *Proceedings of the 23rd ACM SIGPLAN Symposium on Principles and Practice of Parallel Programming*, Vienna Austria, Feb. 2018, pp. 246–260, doi: 10.1145/3178487.3178506.

[10] L. Nai, Y. Xia, I. G. Tanase, H. Kim, and C. Y. Lin, "GraphBIG: understanding graph computing in the context of industrial solutions," in *SC15: International Conference for High Performance Computing, Networking, Storage and Analysis*, Nov. 2015, pp. 1–12, doi:

10.1145/2807591.2807626.

[11] R. C. Murphy, K. B. Wheeler, B. W. Barrett, and J. A. Ang, "Introducing the Graph 500," p. 5.

[12] S. Beamer, K. Asanović, and D. Patterson, "Direction-Optimizing Breadth-First Search," *Sci. Program.*, vol. 21, no. 3–4, pp. 137–148, 2013, doi: 10.1155/2013/702694.

[13] Y. Yasui and K. Fujisawa, "Fast and scalable NUMA-based thread parallel breadth-first search," in *2015 International Conference on High Performance Computing & Simulation (HPCS)*, Amsterdam, Netherlands, Jul. 2015, pp. 377–385, doi: 10.1109/HPCSim.2015.7237065.

[14] C. Zhang, H. Cao, X. Ye, G. Wang, Q. Hao, and D. Fan, "Highly Efficient Breadth-First Search on CPU-Based Single-Node System," in *2019 IEEE 21st International Conference on High Performance Computing and Communications; IEEE 17th International Conference on Smart City; IEEE 5th International Conference on Data Science and Systems (HPCC/SmartCity/DSS)*, Aug. 2019, pp. 2066–2071, doi: 10.1109/HPCC/SmartCity/DSS.2019.00286.

[15] Zhang C. *et al.*, "Efficient Optimization of Graph Computing on High-Throughput Computer," 计算机研究与发展, vol. 57, no. 6, p. 1152, Jun. 2020, doi: 10.7544/issn1000-1239.2020.20200115.

[16] D. Fan *et al.*, "SmarCo: An Efficient Many-Core Processor for High-Throughput Applications in Datacenters," in *2018 IEEE International Symposium on High Performance Computer Architecture (HPCA)*, Feb. 2018, pp. 596–607, doi: 10.1109/HPCA.2018.00057.

[17] M. Faloutsos, P. Faloutsos, and C. Faloutsos, "On power-law relationships of the Internet topology," *ACM SIGCOMM Comput. Commun. Rev.*, vol. 29, no. 4, pp. 251–262, Aug. 1999, doi: 10.1145/316194.316229.

[18] J. Leskovec, D. Chakrabarti, J. Kleinberg, C. Faloutsos, and Z. Ghahramani, "Kronecker Graphs: An Approach to Modeling Networks," p. 58.

[19] "November 2019 Green | Graph 500." https://graph500.org/?page_id=793 (accessed Sep. 06, 2020).